# Real-Time Outlier Connections Detection in Databases Network Traffic

Leonid Rodniansky

IBM Security Guardium Development, lrodnian@us.ibm.com

Tania Butovsky

IBM Security Guardium Development, tbutovsk@us.ibm.com

Mikhail Shpak

IBM Security Guardium Development, Mikhail.Shpak@ibm.com

**Abstract**

The article describes a practical method for detecting outlier database connections in real-time. Outlier connections are detected with a specified level of confidence. The method is based on generalized security rules and a simple but effective real-time machine learning mechanism. The described method is non-intrusive to the database and does not depend on the type of database. The method is used to proactively control access even before database connection is established, minimize false positives, and maintain the required response speed to detected database connection outliers. The capabilities of the system are demonstrated with several examples of outliers in real-world scenarios.

## 1 INTRODUCTION

Real-time outlier detection in database network traffic is important for database security. It allows immediate detection of potential security threats and enables proactive measures to prevent them.

While the field of outlier detection has been thoroughly studied in general, and machine learning methods such as Local Outlier Factor (LOF) [1], DBSCAN [2], Bayesian Networks [3], Isolation Forest [4], [7] etc., greatly help in detecting abnormal behavior, real-time outlier detection in database traffic imposes limitations, and the direct use of standard machine learning methods is problematic.

The specificity of database network traffic implies large volumes of data and database connections per unit of time. This requires machine learning methods to be unsupervised and highly efficient in terms of speed, CPU, and memory usage. Minimizing the response time to a detected outlier is essential for the security of the protected system.

In addition to the speed of detection, high confidence in decisions is important. Access control to the protected system becomes unreliable with false positives. When abnormal connections to a database are detected, the security system can issue alerts, quarantine offenders, or block suspicious connections. False positives can overload Security Information and Event Management (SIEM) systems and disrupt normal operations. Therefore, an important aspect of outlier detection is the ability to optimally distinguish between the learning and detection phases.

This article describes an implementation of outlier detection using the IBM Guardium Data Protection (GDP) [5] system as an example, which is designed to provide access control and alerting in real time without interfering with the database or using direct access to databases.

The paper presents:

- A generalization of security policies for detecting outliers in database connections. This generalization allows specifying the types of outliers that the machine learning engine will look for.
- Several examples of outliers in database connections.
- A simple and effective machine learning engine used to find outlier database connections in real-time.
- Demonstration of alert in real environment.

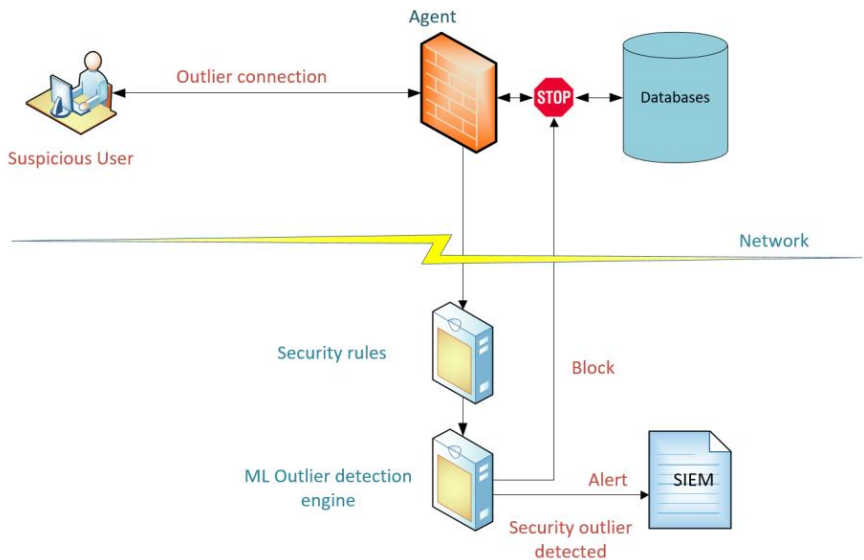

*Figure 1. Blocking and alerting in real-time using outlier detection*

## 2   SYSTEM DETAILS

### 2.1   SECURITY RULES FOR OUTLIER DETECTION

IBM Security Guardium Data Protection's Policy Builder [5] for data is a tool designed to define security policies using customizable rules. These policies ensure robust data protection and can be tailored to meet specific security requirements.

The Policy Builder is accessible via the user interface (UI). Users with appropriate permissions can log in to GDP and access the Policy Builder from the dashboard.

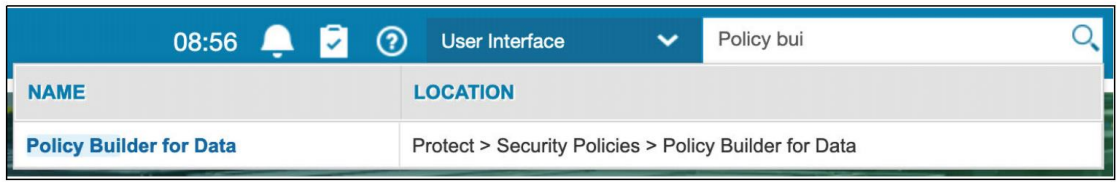

*Figure 2. Starting Policy Builder for Data*

#### 2.1.1 Defining and Detecting Outlier Connections Through Security Rules

Outlier connections to detect are defined through the creation of security policies and their security rules. These rules validate criteria and enforce actions based on specified conditions. This approach provides flexibility, enabling the same detection mechanism to address diverse types of outliers.

Below are several examples of action templates for generating an alert and terminating a suspicious connection.

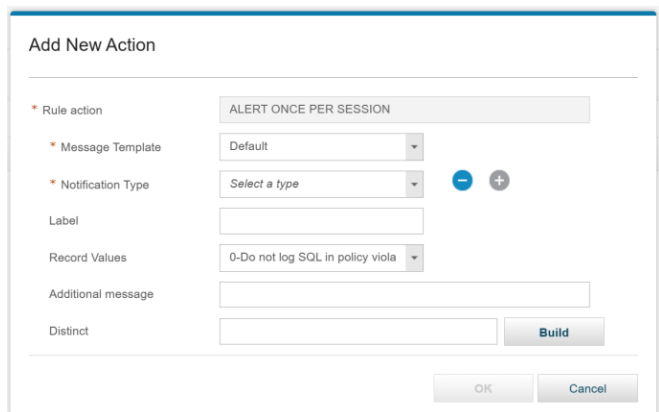

*Figure 3. Action ALERT template*

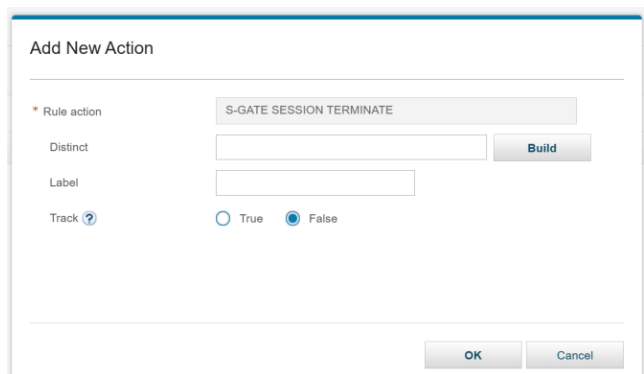

*Figure 4. Action TERMINATE template*

Flexibility is achieved through the use of security rules variables, the DISTINCT parameter, and the OUTLIER operator. The OUTLIER operator, as a component of the DISTINCT parameter, operates on security rule variables. Together, these elements enable the parameterization of security actions, such as generating an ALERT for violations or TERMINATE suspicious connections.

### 2.1.2 Security Rules Variables

Security rule variables are placeholders holding current values for key parameters like database usernames or IPs addresses etc, allowing generalized rules to adapt dynamically to real-time activities. Similar to programming variables, they provide the level of abstraction.

Examples of security rules variables are:

1. Database-specific variables:

   - $(DB_USER)$        – Database user name
   - $(DB_NAME)$        – Database name
   - $(DB_TYPE)$        – Database type
   - $(SERVICE_NAME)$        – Service name

2. Client-specific variables:

   - $(CLIENT_IP)$        – Database client IP address
   - $(CLIENT_HOST_NAME)$        – Client host name
   - $(CLIENT_OS_NAME)$        – Client operating system name
   - $(AUTH_TYPE)$        – Authentication type
   - $(SOURCE_PROGRAM)$        – Application program name
   - $(NET_PROTOCOL)$        – Network protocol
   - $(OS_USER)$        – Operations system user name
   - $(SESSION_INFO)$        – Database session information
   - $(SESSION_KEY)$        – Unique database session key
   - $(CTIMEZONE)$        – Client time zone
   - $(DATETIME)$        – Date and time

3. Server-specific variables:

- $(SERVER_IP)$       – Database server IP address
- $(SERVER_HOST_NAME)$       – Server host name
- $(SERVER_DESC)$       – Server description
- $(SERVER_OS_NAME)$       – Server operating system name
- $(SENDER_IP)$       – Sender's IP address

4. Query-specific variables:

- $(STATEMENT_KEY)$       – Unique database query key
- $(COMMAND)$       – Command issued
- $(ERROR)$       – Error code
- $(CONSTRUCT_KEY)$       – Query construct key
- $(LITERALS_KEY)$       – Query literals key

### 2.1.3 Action Parameter Distinct

The action parameter DISTINCT enables the definition of outliers. In the following imaginary example, an outlier is identified based on an abnormal combination of security rules variables. The machine learning engine will look for outliers where a database user accesses a specific database on a particular server.

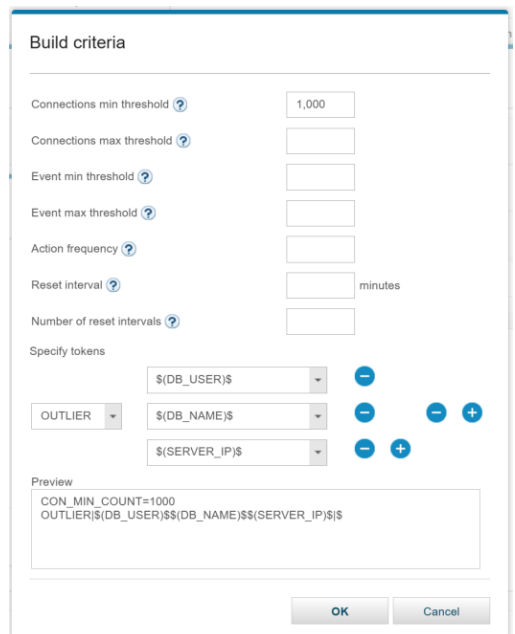

*Figure 5. Outlier definition example*

For large database servers with multiple connections per unit of time, a minimum threshold CON_MIN_COUNT of 1000 observed connections is set to ensure that the learning phase does not finish prematurely. However, further decision to conclude the learning phase is ultimately determined by the machine learning engine and does not depend on this threshold parameter.

### 2.1.3.1    Example 1: Abnormal Authentication type

As an example, suppose database clients are typically configured to authenticate users via Kerberos. Suddenly, a database client attempts direct database authentication using database user credentials. If the machine learning engine is in the outlier detection state, then an alert will be triggered according to the following generic definition:

*Figure 6. Example of abnormal authentication type*

and prepared according to this definition action:

*Figure 7. Prepared action ALERT*

The alert will also trigger in cases of:
- Previously unknown client host name
- Attempts to access a database type not used before by the client

This is a generic alert for outliers of any authentication method, database type and client host name.

### 2.1.3.2    Example 2: Abnormal Time Zone Connection

Another example of an outlier is a database client suddenly connecting from an unusual time zone. Machine learning mechanism in the outlier detection state identifies outlier. Such a connection will not be allowed. Rule definition is:

*Figure 8. Example of abnormal time zone*

and prepared according to this definition action:

*Figure 9. Prepared action TERMINATE connection*

## 2.2   REAL-TIME MACHINE LEARNING FOR OUTLIER DETECTION

*2.2.1 Database Connections*

Database connections are considered as vectors of parameters:

$$C_i = \{c_{i1}, c_{i2}, \dots c_{ik}, \dots c_{im}\}, \quad k \in [1, m], \;\; i \in [1, n] \quad,$$

and we can highlight a subset of connection parameters that are interesting from a security perspective. These subsets of parameters can be hashed:

$$H_i = H(c_{ij1}, c_{ij2}, \dots, c_{ijk}), \quad i \in [1, n], \quad j_1, j_2, \dots, j_k \in [1, m]$$

and the resulting hashes $H_i$ serve as a unique identifier for each connection.

The machine learning mechanism collects and stores these hashes during the learning phase, building a set of unique hashes:

$$S = \{H_1, \dots, H_i, \dots H_n\}, \quad i \in [1, n]$$

that represent the normal baseline of database connections. Once the learning phase concludes, the system uses this baseline to evaluate incoming connections. If a new connection's hash $H_l$ is not found in the set S of established baseline hashes:

$$H_l \notin S, \quad l > n$$

the system declares the connection $H_i$ as an outlier.

This process enables the identification of unusual or potentially malicious connections in real time.

In practice, set S can be implemented as an ordered vector container. Suppose it contains 10,000 hashes, each with a size of 64 bits (8 bytes). Considering the vector's header size of 24 bytes, the total memory required is only:

$$8 * 10000 + 24 = 80024 \; bytes$$

*2.2.2 End of Learning Phase*

Determining when the learning phase has ended is a critical aspect of the machine learning process. If the learning phase concludes too early, it may result in false positives; conversely, if it extends too long, false negatives could occur, leading to potentially malicious database connections going undetected.

Intuitively, if the set of all hash values is finite, then over time, as we observe incoming connections, the number of new, previously unseen connections will decrease. In such cases, we can estimate the probability that all connections have already been observed. If this probability is sufficiently high, the learning phase can be considered complete.

Since the number of users with access to sensitive information on the database server, along with the number of possible applications, operating system users, and similar entities, is finite, and large systems process many connections per unit of time, it is possible to determine with confidence when the learning phase is complete.

Suppose there are $n$ possible independent and distinct connections $H_i$ in total. These connections appear randomly with probability $p_i$, and we have observed them $N$ times ($N > n$). The probability $P_{n,N}$ that all connections have occurred at least once after $N$ observations is:

$$P_{n,N} \geq 1 - \sum_{i=1}^{n} (1 - p_i)^N \tag{1}$$

Let confidence level be $1 - \delta$ where $\delta$ is the small value close to zero. The condition for all $n$ distinct connections to have occurred at least once after $N$ observations, with a confidence level of $1 - \delta$ is:

$$1 - \sum_{i=1}^{n} (1 - p_i)^N \geq 1 - \delta \quad \Rightarrow \sum_{i=1}^{n} (1 - p_i)^N \leq \delta \tag{2}$$

with small $p_i \ll 1$ and large $N$:

$$(1 - p_i)^N < e^{-p_i N}$$

and inequality (2) can be transformed into:

$$\sum_{i=1}^{n} e^{-p_i N} \leq \delta \tag{3}$$

According to *Jensen inequality* for convex functions:

$$\sum_{i=1}^{n} e^{-p_i N} \geq n e^{-\frac{N}{n}} \leq \delta \tag{4}$$

and we have an inequality for the minimal number of required observations $N$ before which there is no point in finishing the training phase [12]:

$$N > n * ln\frac{n}{\delta} \tag{5}$$

This inequality can be also attributed to the statistical "coupon collector problem" [8],[9],[10],[11].

If inequality (5) is true for some values of $n$ (the number of distinct observed hashes) and $N$ (the total number of observed hashes), then this means that with probability approaching $1 - \delta$, the security system has observed all possible distinct connections $H_i$.

Ideally, if the probabilities of the occurrence of distinct connections are equal $p_i = \frac{1}{n}, \forall i \in [1, n]$, then the probability of observing all connections is exactly $1 - \delta$.

During security system monitoring, note that both $n$ and $N$ can increase over time. Also, inequality (5) was derived without the assumption that the probabilities $p_i$ form a complete probability distribution, i.e.:

$$\sum_{i=1}^{n} p_i \leq 1$$

The learning phase is considered complete when equality (5) is satisfied.

Inequality (5) in graphical form for confidence 95% (δ=0.05):

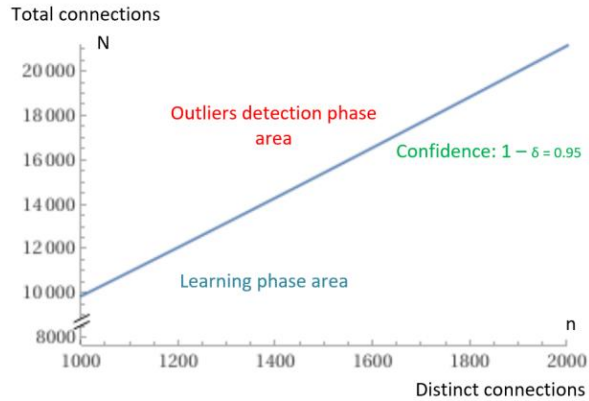

*Figure 10. Outlier detection and learning phases. Confidence 95% (δ=0.05).*

When $\delta = 0.05$, a new, previously unseen connection with:

$$H_l \notin S, \; l > n$$

received while the machine learning engine is in the *outlier's detection phase* will be considered an outlier with a confidence level close to 95%.

In practice, however, outliers may be caused by environmental changes, such as the addition of new applications, new groups of users, or other operational modifications. Upon identifying and alerting such instances, the system updates the set of hashes $S$ to incorporate these changes. After re-evaluating inequality (5) to ensure it aligns with the updated data, the system may revert to the learning phase to accommodate and integrate the new connection patterns into its baseline.

# 3 IMPLEMENTATION DETAILS

Security rules UI was implemented in Java [5], and the core security rules engine is written in C++. The machine learning algorithm has been implemented in C++. Partial code shown in simplified Algorithm 1 uses the C++ standard libraries <algorithm> and <vector>. Hashes $H_i$ are calculated using 64-bit MurmurHash3 [6]. The hashes container is implemented as an ordered vector. The use of an ordered vector ensures memory efficiency and O(log(n)) complexity for lookups after sorting. Memory is proportional to the number of unique hashes stored in the container.

---

### ALGORITHM 1: Machine Learning Method

Initially set security rule 'learning phase', N = 0, n = 0, δ = 1 - confidence probability

A1: For each incoming connection:

    Compute hash $H_i$ for group of connection parameters defined in security rule.

    N = N + 1

    If $H_i$ was not within ordered vector container

        n = n +1

        Insert $H_i$ into ordered vector container

        If 'detection phase'

            Issue alert or block connection $H_i$

    Check inequality (5)

        If 'true' then

            Set 'detection phase'

        Else

            Set 'learning phase'

End A1

---

Different security rules are checked in parallel in a multi-threaded process. Ordered vectors allow parallel searches. This means that multiple threads can simultaneously check security rules and detect outliers for different database connections independently of each other.

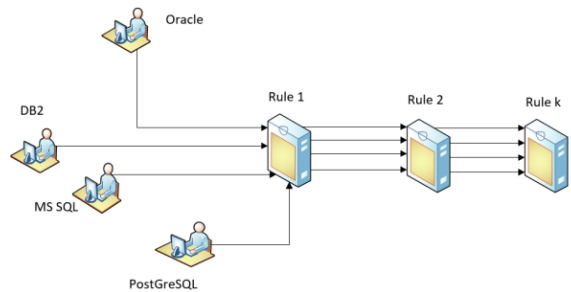

*Figure 11. Parallelization of verification of heterogeneous database connections*

# 4  RESULTS

One setup included an Oracle database server running on a Red Hat Enterprise Linux 9 server with 4 physical CPUs, 14 cores per CPU, and 128 GB of memory. The number of database users was 120 and the number of OS users was 30. Some OS users were prohibited from initiating connections with certain database users.

Database clients connected randomly using non-interactive JDBC connections to the database server. A security rule set to look for database users connecting as unexpected OS users. The confidence level is set to 95% to avoid false positives.

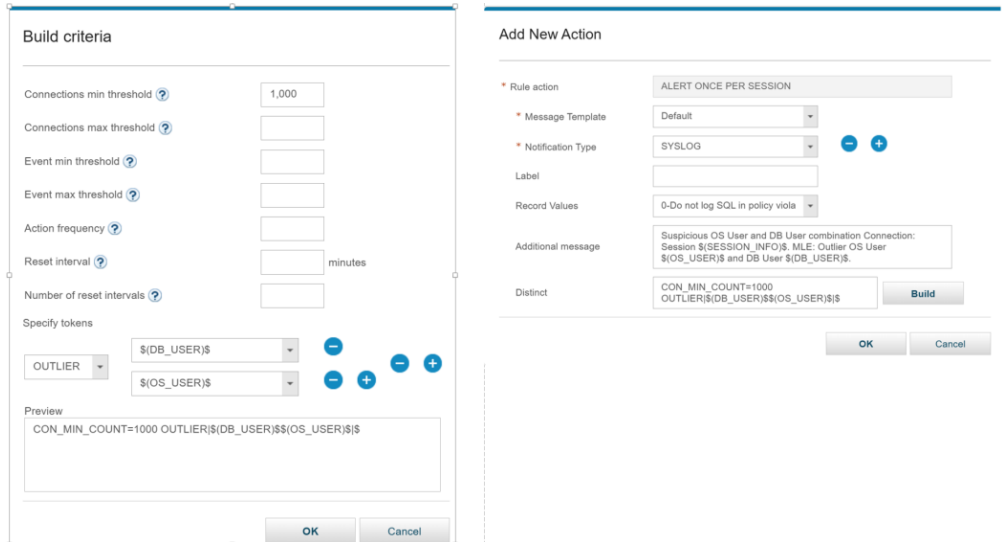

*Figure 12. Generalized action definition*

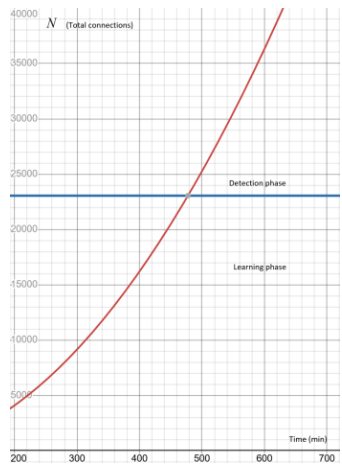

*Figure 13. Connections rate*

After ~8 hours of observation, the security system switched to the outlier detection phase. At this moment, $n = 2160$ distinct connections $H_l$ were observed from, $N = 23100$ in total. According to inequality (5), the security system, with a high level of confidence, observed all possible valid distinct connections. As a result, the security system entered the detection phase, did not issue any alerts, and the number of distinct connections remained stable. At some point during monitoring, the security system generated an alert when an unexpected administrative OS user connected to the database server using valid database user credentials.

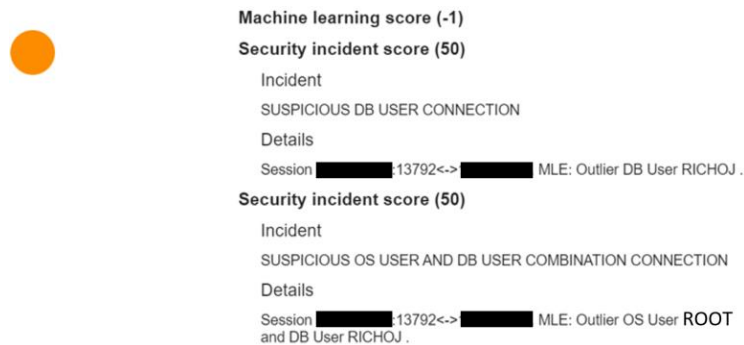

*Figure 14. Outlier OS user/DB user connection*

The total overhead for response time was about 5 milliseconds, considering the transfer of information over the network. Such a delay is completely acceptable when establishing connections to the database.

## 5 CONCLUSION

The paper discussed a practical method for identifying database connection outliers using generalized security policies and efficient real-time machine learning method.

Key characteristics of the method include:

- Controllable level of confidence in real-time outlier detection and reasonable switching between training and detection phases of the outlier detection mechanism minimize false positives.
- Independence from database types, which allows using the same outlier detection security policies and mechanisms in different environments.
- Flexibility of security policies and their rules, allowing to determine the types of outliers to detect.
- Relatively low memory footprint of the real-time machine learning mechanism.
- Parallelization of outlier detection for simultaneous heterogeneous database connections.
- High response speed to outlier detection (complexity O(log n)), which ensures effective access control.
- No interference in database operations or direct access to databases.
- Outlier database connections can be blocked before they are established.

# REFERENCES

[1]  Breunig, M. M.; Kriegel, H.-P.; Ng, R. T.; Sander, J.. *LOF: Identifying Density-based Local Outliers* (PDF). *Proceedings of the 2000 ACM SIGMOD International Conference on Management of Data*. SIGMOD. pp. 93–104. doi:10.1145/335191.335388. ISBN 1-58113-217-4, 2000

[2]  Ester, Martin; Kriegel, Hans-Peter; Sander, Jörg; Xu, Xiaowei. Simonds, Evangelos; Han, Jiawei; Fayyad, Usama M. (eds.). *A density-based algorithm for discovering clusters in large spatial databases with noise*, 1996

[3]  Ethan Roberts, Bruce A. Bassett, Michelle Lochner. Bayesian Anomaly Detection and Classification. 2019, https://arxiv.org/abs/1902.08627.

[4]  *Liu, Fei Tony; Ting, Kai Ming; Zhou, Zhi-Hua, "Isolation Forest". 2008 Eighth IEEE International Conference on Data Mining. pp. 413–422. doi:10.1109/ICDM.2008.17. ISBN 978-0-7695-3502-9. S2CID 6505449,* 2008

[5]  IBM Guardium. IBM Guardium Data Protection v.12.1, Security anomalies, 2024

[6]  Austin Appleby, "SMHasher", Github.com, 2016

[7]  Yuanyuan Luo; Xuhui Du; Yi Sun. Survey on Real-time Anomaly Detection Technology for Big Data Streams. 2018 12[th] IEEE International conference, 2018

[8]  Rajeev Motwani, Prabhakar Raghavan, Randomized Algorithms. Cambridge University Press, 0-521-47465-5, 1995

[9]  "Coupon collector's problem", Wikipedia, The Free Encyclopedia, 2024, https://en.wikipedia.org/wiki/Coupon_collector%27s_problem

[10]  Wenyu Xu, A. Kevin Tang, A Generalized coupon collection problem. J. Appl. Prob. 48, 1081–1094. 2011

[11]  S. N. Bernstein, The Theory of Probabilities (Russian), Moscow, Leningrad, 1946

[12]  L, Rodniansky, T. Butovsky, M. Shpak, "Identifying outlier application connections to services with controlled confidence level and in real-time ", Patent application US20240106860A1. 2022